\documentclass[rnote]{aa} % for the research notes
\usepackage{graphicx}
\usepackage{txfonts}
\usepackage{caption}
\usepackage{subcaption}
\begin{document}

   \title{NGC 5523: An Isolated Product of Soft Galaxy Mergers?}

   \author{Leah M. Fulmer
          \inst{1}, John S. Gallagher, III
          \inst{1}
          \and
          Ralf Kotulla\inst{1}}

   \institute{Department of Astronomy, University of Wisconsin - Madison, WI, USA 53706\\
%              \email{lfulmer@wisc.edu}
             }

% \abstract{}{}{}{}{} 
% 5 {} token are mandatory
 
  \abstract
{Multi-band images of the very isolated spiral galaxy NGC 5523 show a number of unusual features consistent with NGC 5523 having experienced a significant merger: (1) Near-infrared (NIR) images from the Spitzer Space Telescope (SST) and the WIYN 3.5-m telescope reveal a nucleated bulge-like structure embedded in a spiral disk. (2) The bulge is offset by $\sim$1.8 kpc from a brightness minimum at the center of the optically bright inner disk. (3) A tidal stream, possibly associated with an ongoing satellite interaction, extends from the nucleated bulge along the disk. We interpret these properties as the results of one or more non-disruptive mergers between NGC 5523 and companion galaxies or satellites, raising the possibility that some galaxies become isolated because they have merged with former companions.$^0$}


   \maketitle
%
%________________________________________________________________

\section{Introduction}

NGC 5523 is an apparently normal late-type isolated spiral galaxy (\cite{kar11}; \cite{min16}) with structural properties indicative of galaxy-galaxy interactions. While isolated galaxies are often considered to evolve with minimal interactions from surrounding galaxies (\cite{jnc09}), we show that NGC 5523 offers a potential counter-example to this assumption. \footnotetext{The data used in Figures 1, 3, and 4 are available in electronic form
at the CDS via anonymous ftp to cdsarc.u-strasbg.fr (130.79.128.5)
or via http://cdsweb.u-strasbg.fr/cgi-bin/qcat?J/A+A/}

\cite{ver05} defined isolated galaxies as those that have not been affected by surrounding objects in the past 3 Gyr. More recently, \cite{kar11} shortened this history of interaction, defining isolation as a lack of interaction in the past 1-2 Gyr. By either definition, the structural properties of isolated galaxies are typically assumed to have evolved almost exclusively in their initial formation. Galaxies may be isolated individually or isolated in pairs, triplets, and groups. Of course, in the latter cases, the entire group has not experienced an external influence in the said length of time. In the case of NGC 5523, its nearest neighbor lies more than 1 Mpc away, suggesting that it has not interacted with any of its current neighbors for at least several Gyrs. However, its asymmetric internal features imply recent interactions, raising the possibility that it may have become isolated in conjunction with evolutionary processes.

Asymmetries in galaxies can result from a variety of processes, including gas accretion, fly-by interactions, and minor mergers (e.g., \cite{znr97}; \cite{bjc05}; \cite{map08}). Furthermore, once formed, asymmetries may persist over Gyr time scales (\cite{znr97}; \cite{scj07}; \cite{lot15}). Thus, a disturbed structure in a galaxy can result from processes that occurred sufficiently long ago that obvious remnants, such as the interacting companion, no longer exist.

A detailed study of NGC 5523 has been conducted in which \cite{min16} conclude that NGC 5523 shows no evidence for recent interactions based on its Hydrogen I (HI) line profile. Minchin et al. reach this conclusion by investigating companion candidates near NGC 5523 and ultimately find that no companions exist within 300 kpc of NGC 5523. This result is based on HI data in the ALFALFA survey. 

We show that NGC 5523 has properties associated with a historic merger, has experienced a minor interaction in the recent past, and has two dwarf galaxy companions within a 2 Mpc radius. Our study is based on a combination of WIYN H-band imaging, Sloane Digital Sky Survey (SDSS) optical data, and Spitzer Space Telescope (SST) Infrared Array Camera (IRAC) 3.6 $\mu$m data. Section 2 describes the observations and data reduction processes that provide the basis for our research. Section 3 details the significant features that lead us to believe that NGC 5523 has participated in past mergers. Section 4 discusses the morphological implications of NGC 5523's abnormal internal structure. Finally, Section 5 offers our conclusions concerning the past of this particular galaxy and what this research could imply for other isolated galaxies.
\begin{table}[h!]
\caption{A general overview of NGC 5523.}             % title of Table
\label{table:1}      % is used to refer this table in the text
\centering                          % used for centering table
\begin{tabular}{l l}        % centered columns (4 columns)
\hline                 % inserts double horizontal lines
\noalign{\smallskip}
Property & Data \\    % table heading 
%\noalign{\smallskip}
\hline                        % inserts single horizontal line
\noalign{\smallskip}
Right Ascension &14:14:52.3 (J2000)\\
Declination &25:19:03 (J2000)\\
Distance &20.6 Mpc\\
Redshift &1039  km/s\\
Classification &SA(s)cd\\
Rotational Velocity &150 km/s\\
Stellar Mass &$\sim$1$\times 10^{10}$ M$_\odot$\\
\hline                                   %inserts single line
\end{tabular}
\end{table}
\tablefoot{The NASA/IPAC Extragalactic Database (NED) is the source of all information in Table 1, except if noted otherwise. The rotational velocity was calculated from the H1 line profile. The stellar mass was computed from a photometric analysis of a sky-subtracted 3.6$\mu$m SST image of NGC 5523.}

\section{Observations and data reduction}
\subsection{Observational data}

Our multiwavelength observations include images from Spitzer Space Telescope, Sloane Digital Sky Survey, and WIYN. A sky-subtracted version of the archival near-infrared (NIR) images from SST (IRAC) at 3.6$\mu$m and $\sim$2\textquotedblright \ angular resolution\footnote{Data Set ID: 'ads/sa.spitzer$\#$0030670592'} (PI: Kartik Sheth, Sky-subtraction: Liese van Zee) were used to capture NGC 5523's internal structures, including the tidal stream. The combination of SST and SDSS images allow us to identify an offset of $\sim$1.8 kpc between the centers of the optically bright inner disk and the larger outer disk. 

The NIR H-band image was obtained on May 24, 2010 using the WIYN High Resolution Infrared Camera (WHIRC) on the WIYN 3.5-m telescope (angular resolution $\sim$0.8\arcsec) located on Kitt Peak\footnote{Data Set ID: 'wiyn.whirc.20100525T082757'} (PI: Ralph Kotulla). We observed NGC 5523 using a four-point dither pattern with 40\arcsec steps, repeated twice. A separate sky field, offset by 5\arcmin, was observed between the two repeats. Individual exposure times for both object and sky pointings were 100 seconds per exposure for a total on-source integration time of 800s. Data reduction followed the WHIRC Data Reduction manual: All exposures were initially corrected for non-linearity using the coefficients in the data reduction guide. Flat-fields were computed from the difference between ten dome-flat exposures with flat-field lamps turned on and off, and corrected for pupil ghost contribution using the published WHIRC pupil ghost template. Each frame was then sky-subtracted using a median-combined stack of all sky-field exposures, and flat-fielded using the aforementioned flat-field. All frames were then distortion-corrected and co-aligned by matching the position of all stars detected in each field, and finally combined into the final stack shown in Figure 5. Astrometric and photometric calibration was achieved by matching all point sources detected in the stack to a catalog of sources from the Two Micron All-Sky Survery (2MASS) Point Source Catalog (PSC, Skrutskie et al. 2006), providing a matched source catalog of seven stars and yielding absolute astrometric errors of $\pm 0.2$\arcsec and an absolute photometric calibration uncertainty of $\sim 0.15$ mag.

\subsection{Isophotal modeling and magnitudes}
The Space Telescope Science Data Analysis System (STSDAS) modeling task {\tt Ellipse} was used on the SST 3.6$\mu$m image in order to measure radial light profiles and examine the structure of a possible tidal stream in contrast with the inner disk isophotes. We first performed three individual {\tt Ellipse} analyses in order to note the changes in the model symmetry depending on the given center point. Table 2 describes the consistent parameters between the three analyses, and Table 3 describes the unique parameters of each analysis with regard to the chosen center point and parameter fixedness. Figure 1 shows the three {\tt Ellipse} analyses individually overplotted onto a gray-scale image of the input sky-subtracted 3.6$\mu$m SST image.
\begin{table}[h!]
\caption{Consistent input parameters over all {\tt Ellipse} analyses of NGC 5523.}             % title of Table
\label{table:2}      % is used to refer this table in the text
\centering                          % used for centering table
\begin{tabular}{l l}        % centered columns (4 columns)
\hline                 % inserts double horizontal lines
\noalign{\smallskip}
Parameter & Data \\    % table heading 
%\noalign{\smallskip}
\hline                        % inserts single horizontal line
\noalign{\smallskip}
Input Image  &Sky-Subtracted 3.6$\mu$m SST Image\\
Isophote Sample & Median\\
Max. Semi-Major Axis & 13.2 kpc\\
Brightness Max. Coordinates &(14:42:52.4, +25:19:03.6)\\
Brightness Min. Coordinates &(14:14:51.6, +25:19:05)\\
\hline                                   %inserts single line
\end{tabular}
\end{table}

\begin{table}[h!]
\caption{Variable parameters among {\tt Ellipse} analyses to account for offset of centers.}             % title of Table
\label{table:3}      % is used to refer this table in the text
\centering                          % used for centering table
\begin{tabular}{l l l l}        % centered columns (4 columns)
\hline                 % inserts double horizontal lines
\noalign{\smallskip}
Figure & Center & Initial Ellipticity & Initial PA\\    % table heading 
%\noalign{\smallskip}
\hline                        % inserts single horizontal line
\noalign{\smallskip}
1(a) & Brightness Max, Fixed  & Unfixed & Unfixed\\
1(b) & Brightness Min, Fixed  & Fixed & Fixed\\
1(c) & Brightness Max, Unfixed  & Unfixed & Unfixed\\
\hline                                   %inserts single line
\end{tabular}
\tablefoot{We tried to perform an analysis with the parameters Brightness Min, Fixed / Unfixed / Unfixed, but {\tt Ellipse} could not maintain a logical continuation of the analysis given these parameters.}
\end{table}
\begin{figure}[htp]

\centering
\includegraphics[width=.225\textwidth,angle=270]{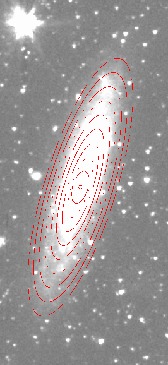}
\includegraphics[width=.2372\textwidth,angle=270]{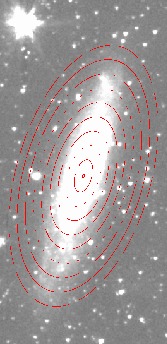}
\includegraphics[width=.225\textwidth,angle=270]{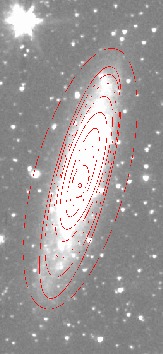}

\caption{{\tt Ellipse} models of NGC 5523 overlayed on their original input image, a sky-subtracted version of the archival 3.6$\mu$m SST image. The top-most figure is described as Figure 1(a) in Table 3, the middle figure as 1(b), and the bottom-most figure as 1(c). All parts of Figure 1 are rotated clockwise 45$^{\circ}$ as compared to Figure 2(a).}
\label{fig:figure1}

\end{figure}
Following these analyses, we used the resulting models to create radial light profiles of NGC 5523. Figure 3 contains the light profile of Figure 1(c), which most obviously displays the offset of centers between the inner and outer disks, represented as an asymmetry in {\tt Ellipse}. This offset of centers is discussed in more detail in Section 3.3.

The task {\tt Apphot} was then used to determine physical properties of the tidal stream and the nucleated bulge, such as luminosity and mass, from the 3.6~$\mu$m image obtained with SST (Table 4). With these mass calculations, we could then determine with more accuracy the influence of such a massive object on NGC 5523's evolution. Individual features of NGC 5523 are detailed in Section 3.

\begin{table}[h!]
\caption{A photometric evaluation of stellar masses for the significant features of NGC 5523.}             % title of Table
\label{table:4}      % is used to refer this table in the text
\centering                          % used for centering table
\begin{tabular}{l l l l}        % centered columns (4 columns)
\hline                % inserts double horizontal lines
\noalign{\smallskip}
Object & Area (kpc$^2$) & [3.6] & Mass (M$_\circ$) \\    % table heading 
\hline                        % inserts single horizontal line
\noalign{\smallskip}
Galaxy Total (Outer Disk) &1110.6 &9.5 &$\sim$1.0$\times 10^{10}$\\
Nucleated Bulge &10.3 &12.9 &$\sim$5.5$\times 10^{8}$\\
Visible Tidal Stream &6.9 &17.1 &$\sim$1.2$\times 10^{7}$\\
\hline                                   %inserts single line
\end{tabular}
\tablefoot{All values in Table 4 are derived from {\tt Apphot} analyses of a sky-subtracted 3.6 $\mu$m SST image of NGC 5523. The calculations of stellar masses assume $\frac{M_\circ}{L_\circ}=1.0$. Figure 2 exhibits the contour levels that defined these regions in our photometric analyses.}
\end{table}

\begin{figure}[htp]

\centering
\includegraphics[width=.5\textwidth]{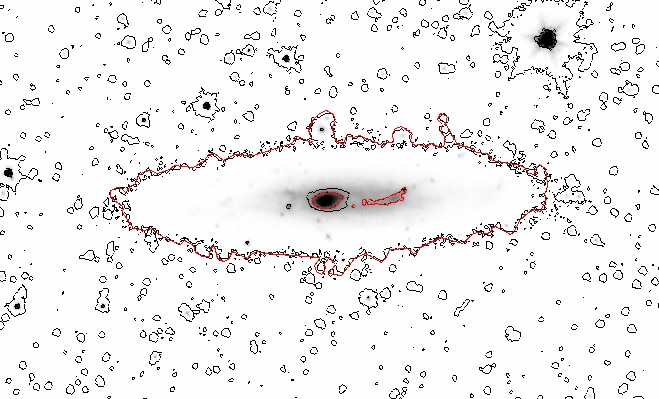}
\includegraphics[width=.5\textwidth]{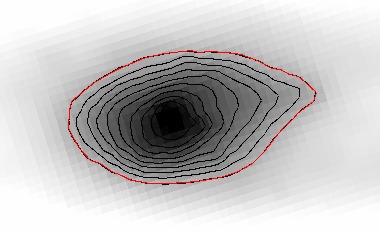}
\includegraphics[width=.5\textwidth]{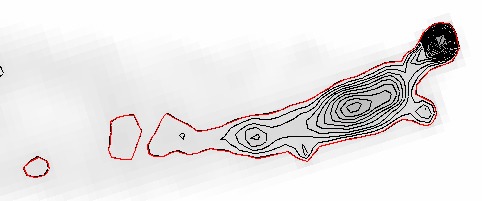}

\caption{Contour levels used for the {\tt Apphot} analyses of the Galaxy Total (top, 2(a)), Nucleated Bulge (middle, 2(b)), and Visible Tidal Stream (bottom, 2(c)) as described in Table 4. All contours were based on a sky-subtracted version of the archival 3.6$\mu$m SST image. For each analysis, the outermost contour (drawn in red) was taken as the sky value, thus {\tt Apphot} considered all light within each exhibited contour in the photometric analysis. The physical dimensions of the topmost image (2(a)) are 44.7 kpc x 30.4 kpc.}
\label{fig:figure2}

\end{figure}

\section{Analysis of significant features}
\subsection{The regular, blue outer disk}

The outer disk of NGC 5523 displays a highly symmetrical HI line profile and extremely regular structure \cite{hay11}. As seen in Figure 3, there are no significant perturbations in the outer disk, giving a smoothly exponential curve in the latter half of the plot. 

\begin{figure}[h!]
    \centering
    \includegraphics[scale=0.33]{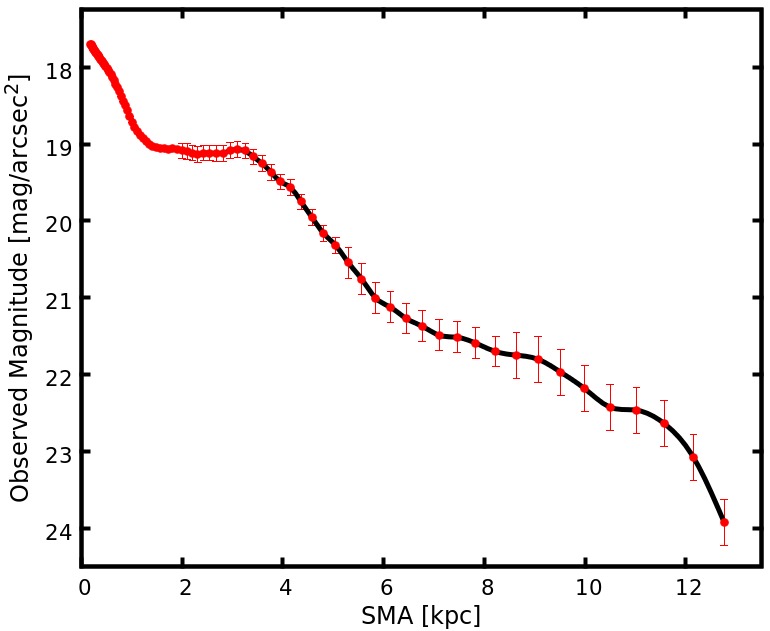}
    \caption{Observed 3.6$\mu$m Magnitude vs. Semi-Major Axis (SMA) for the inner and outer disks of NGC 5523, from the {\tt Ellipse} analysis displayed in Figure 1(c). In this plot, SMA=0 is located at the center of the brightness minimum, and thus the plateau beginning at $\sim$1.8 kpc denotes the presence of the nucleated bulge. Outside the inner disk (starting $\sim$6 kpc), the slope of the outer disk photometry is $\sim$0.34 $\frac{mag}{arcsec^2 \ kpc}$ (displayed as $\sim$-0.34 $\frac{mag}{arcsec^2 \ kpc}$ above). The radial scale length of the outer disk is $\sim$1.4 kpc. The error in magnitude for the inner disk is less than 0.1, improving inward (not shown). Error maximizes at 0.3 in the outer disks ($\sim$3$\sigma$), dominated by systematic background errors. All error measurements are approximate due to uncertainties in the background subtraction.}
    \label{fig:figure3}
\end{figure}

Furthermore, a color analysis of SDSS images of NGC 5523 reveals that the outer disk is relatively blue compared to the nucleated bulge and inner disk (Figure 4). The color analysis was performed by first obtaining {\tt Ellipse} fits for g and r SDSS images. In running the {\tt Ellipse} fits, the SDSS nano-maggy calibration was used to convert to SDSS AB magnitudes. Colors were calculated as the difference between the g and r AB magnitudes from their respective {\tt Ellipse} fits.

\begin{figure}[h!]
    \centering
    \includegraphics[scale=0.33]{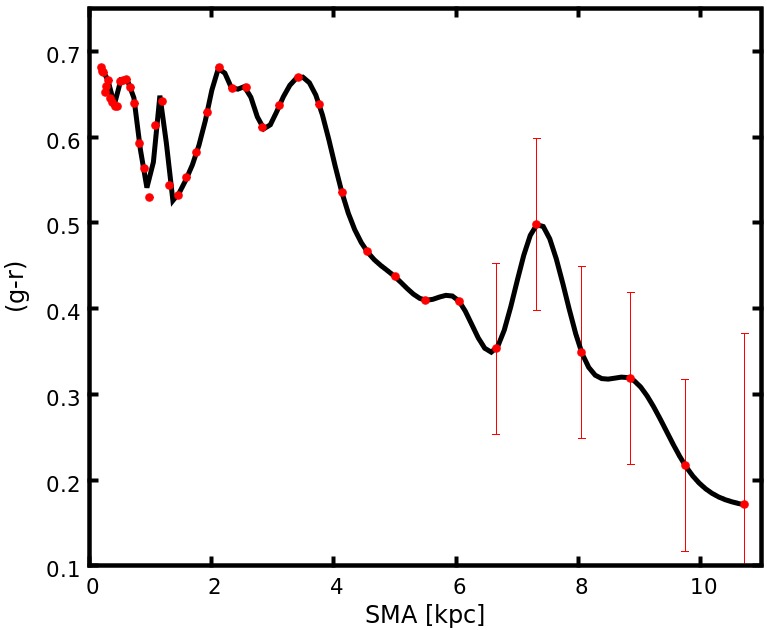}
    \caption{Color analysis of NGC 5523 from SDSS (g-r) images. Color errors for the inner disk are less than 0.1 (not shown). Error maximizes at 0.2 at the outermost isophote, dominated by systematic background errors. All error measurements are approximate due to uncertainties in the background subtraction.}
    \label{fig:figure4}
\end{figure}

\subsection{The bulge and nucleus}

Using WHIRC on the WIYN telescope, we can clearly observe the internal structure of the nucleated bulge, the likely dynamical center of the system (\cite{wal05}). As shown in Figures 2 and 5, the bulge, which produces $\sim$1\% of the 3.6$\mu$ luminosity in NGC 5523 (Table 4), contains a concentrated nuclear feature centered within a significant bulge. This dynamically relaxed structure with the red colors of an old stellar population further suggests that the bulge has long been present and prominent in the host galaxy of this system. The slightly rectangular isophotes surrounding the nucleated bulge in Figure 2 may also be evidence for a bar in the nucleated bulge.

We obtained photometry of the semi-stellar nucleus in NGC 5523 from the H-band WHIRC data. We did not fully resolve the nucleus, but the radial intensity distribution is somewhat broader than that of nearby stars (FWHM seeing of $\sim$0.8-0.9\arcsec). Instrumental magnitudes were calculated with the IRAF {\tt Apphot} task for several circular apertures. Using the magnitude growth curves for stars we adopted results for the nucleus from a seven-pixel radius (1.4\arcsec diameter) aperture. The absolute calibration was obtained by matching the H=10.60 magnitude listed in NED from a 51.8\arcsec diameter aperture on NGC 5523, which gives H=15.8 for the nucleus. For D=20.6 Mpc the nucleus has, coincidentally, M$_H \approx -$15.8. Thus the nuclear complex of NGC~5523 appears to be relatively normal when compared to other spirals (e.g., \cite{fro85}, \cite{hug05}).

\begin{figure}[h!]
    \centering
    \includegraphics[scale=0.412]{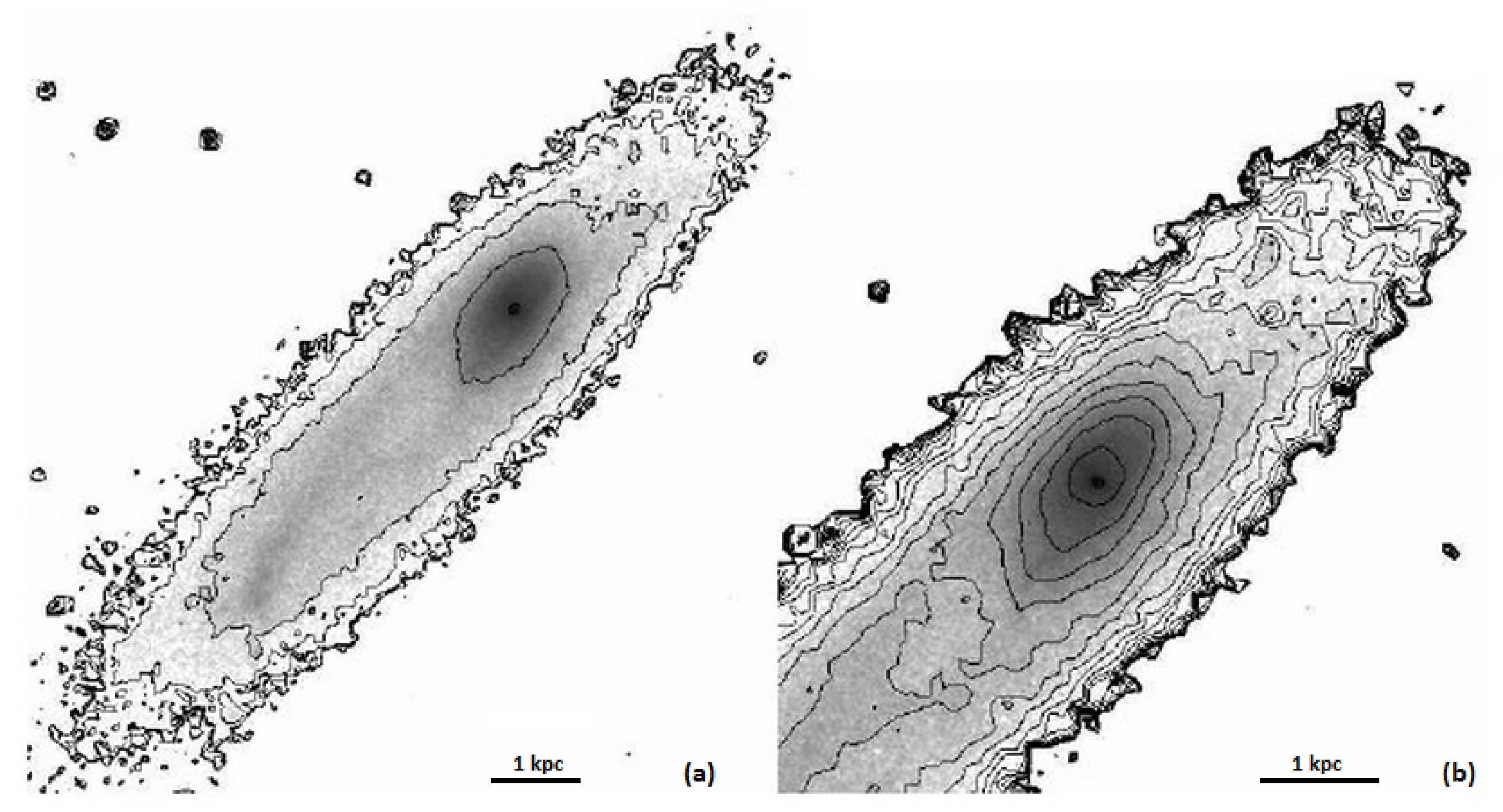}
    \caption{(a) WHIRC H band image from the WIYN 3.5-m telescope, shown in a log scale. WIYN's 0.8\textquotedblright \ resolution allows us to examine the internal structure of the nucleated bulge, offering a level of detail unavailable in IRAC images. (b) The same image with added contours that detail the internal structure of the nucleated bulge. This image also features the brightness minimum in the lower left-hand corner, clearly denoting the offset between the two extrema. Both parts of Figure 5 are rotated counter-clockwise 135$^{\circ}$ as compared to Figure 2(a).}
    \label{fig:figure5}
\end{figure}

\subsection{The offset of centers}

Our surface photometry shows that NGC 5523 consists of two distinct disks: the optically bright inner disk with an intensity minimum at its center and the structurally regular outer disk with the nucleated bulge at its center. These disks, while concentric in typical galaxies, are displaced in NGC 5523. Table 5 gives quantitative evidence for this offset, supplemented with Figure 6. Figure 6 also displays isophotal definitions of the inner and outer disks. As shown in Table 5, the brightness minimum and the nucleated bulge (brightness maximum) are each associated with distinct physical centers, similar to the effects observed in interacting galaxies (\cite{sch96}; \cite{jnm06}). However, asymmetric galaxies are also found in low density environments with no direct indications for recent interactions (e.g., \cite{mng97}, \cite{mng02}).

\begin{table}[!h]
\caption{A quantitative spatial analysis of the offset of centers of NGC 5523.}             % title of Table
\label{table:5}      % is used to refer this table in the text
\centering                          % used for centering table
\begin{tabular}{l l l l}        % centered columns (4 columns)
\hline                 % inserts double horizontal lines
\noalign{\smallskip}
Measuring From & Measuring To & Dist. (kpc) & Diff. (kpc) \\    % table heading 
\hline                        % inserts single horizontal line
\noalign{\smallskip}
\smallskip
Brightness Min &Brightness Max &1.8 &     \\
A &B &24.8 &    \\
Brightness Min &A &10.9 &     \\
Brightness Min &B &13.8 &2.9\\
Brightness Max &A &12.7 &     \\
\smallskip
Brightness Max &B &12.0 &0.7\\
C &D &8.4 &     \\
Brightness Min &C &3.9 &     \\
Brightness Min &D &4.5 &0.6\\
Brightness Max &C &5.7 &     \\
Brightness Max &D &2.7 &3.0\\
\hline                                   %inserts single line
\end{tabular}
\tablefoot{(1) See Figure 6 for the locations of points A, B, C, and D and the locations of the brightness extrema. (2) The Diff. (kpc) values are calculated as the difference between the measurements from each extrema to each apside. This is effectively a quantitative representation of the symmetry of each extrema within each disk. As denoted in Table 5, the difference between the apside measurements of the outer disk is minimized when measured in relation to the Brightness Maximum (nucleated bulge). This means that the bulge lies at the center of the outer disk. Alternatively, the difference between the inner disk apside measurements is minimized when measured in relation to the Brightness Minimum. Thus, the minimum is the center of the inner disk.}
\end{table}   

\begin{figure*}[!ht]
    \centering
    \includegraphics[width=\textwidth]{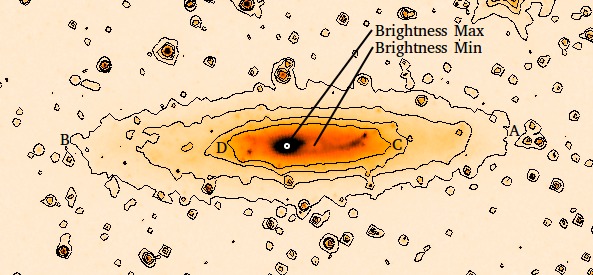}
    \caption{Spitzer IRAC image of NGC 5523 at 3.6$\mu$m, displayed in a log scale. Figure 6 functions as a reference for distance measurements in Table 5. Points A and B denote the apsides of the outer disk, and similarly points C and D denote those of the inner disk. The Brightness Maximum indicates the center of the nucleated bulge and the Brightness Minimum describes the isophotal center of NGC 5523. The major axis runs east-west with east at the right edge of the image and north at the bottom edge.}
    \label{fig:figure6}
\end{figure*}

\subsection{The tidal stream}
We find an asymmetric feature shown in Figure 5 (see also Figure 2a,c) that originates from the nucleated bulge (possibly the associated satellite galaxy) and extends throughout the entire inner disk with an irregular luminosity gradient. This type of structure is usually associated with a tidal stream resulting from the disruption of a dwarf satellite galaxy (e.g., \cite{joh99}, \cite{pen06}).

Figure 5a (see also Figure 6) effectively displays the prominence of the tidal stream, and Table 4 gives a photometric analysis of the visible tidal stream.  The photometry in Table 4 suggests that the tidal stream is associated with an interacting dwarf satellite galaxy with M$_*$ $\geq$ 10$^7$ M$_{\odot}$. The IRAC image shows a bright source at the head of the stream. This feature is not detected in the WIYN H-band image. Therefore, although its nature is uncertain, we can deduce that this feature has very red colors.

\subsection{Neighboring galaxies}

Utilizing NED, we find that NGC 5523 has two notable dwarf galaxy neighbors, as described in Tables 6 and 7. It is unlikely that these companions are gravitationally bound to NGC 5523, making NGC 5523 an at-large member of a loose cloud of galaxies. The presence of these moderately luminous, albeit distant companions suggests that deeper surveys for additional neighboring low-mass galaxies could provide insights into the evolution of NGC 5523.

\begin{table}[h!]
\caption{A summary of the two dwarf galaxy companions of NGC 5523.}             % title of Table
\label{table:6}      % is used to refer this table in the text
%\centering                          % used for centering table
\begin{tabular}{l l l}        % centered columns (4 columns)
\hline                 % inserts double horizontal lines
\noalign{\smallskip}
Object&Separation (Mpc)& Redshift Velocity (km/s)\\    % table heading 
\hline                        % inserts single horizontal line
\noalign{\smallskip}
Companion 1&1.4&966    \\
Companion 2&1.6&1145   \\
\hline                                   %inserts single line
\end{tabular}
\tablefoot{Companion 1: 2MASX J13584501+2409048; Companion 2: CGCG 133-084; NED is the source of all information in Table 6.}
\end{table}

\begin{table}[h!]
\caption{Magnitude and luminosity comparisons of NGC 5523 and its dwarf galaxy companions.}             % title of Table
\label{table:7}      % is used to refer this table in the text
%\centering                          % used for centering table
\begin{tabular}{l l l l}        % centered columns (4 columns)
\hline                 % inserts double horizontal lines
\noalign{\smallskip}
Object&$g$& L$_{opt}$ Ratio&HI Mass (log $M_{\circ}$)\\    % table heading 
\hline                        % inserts single horizontal line
\noalign{\smallskip}
NGC 5523&13.1&1.00&$\sim$9.7    \\
Comp 1&16.8&$\sim$0.10&Not detected in ALFALFA    \\
Comp 2&15.6&$\sim$0.03&$\sim$8.6   \\
\hline                                   %inserts single line
\end{tabular}
\tablefoot{Apparent $g$ for NGC 5523 calculated following Jordi et al. 2006 and adopting B-V=0.59 from NED. The L$_{opt}$ Ratio is the optical luminosity ratio relative to that of NGC 5523.}
\end{table}

\section{Discussion}

NGC 5523 is a globally isolated galaxy (\cite{kar11}; \cite{min16}) in which we find possible evidence for past interactions with neighboring galaxies. This observation contradicts the simplifying assumption that isolated galaxies form and evolve with minimal interactions from other galaxies (\cite{kap05}). Our measurements of the structure of NGC 5523 in combination with results from numerical models (\cite{bjc05}; \cite{map08}; \cite{par16}) lead us to believe that some galaxies that were once part of binary systems or small groups become isolated by way of one or even several minor mergers. This evolutionary path would be consistent with the observed presence of small galaxy groups even in voids where the overall density of galaxies is low (\cite{kre11}).

While asymmetric galaxy structures can result from a variety of processes (gas accretion, fly-by interactions, etc.), we favor a past interaction as the source of the asymmetry in NGC 5523. Gas accretion, for example, is expected to have a larger effect on the HI disk than on the stellar body. The HI spectrum of NGC 5523, however, is highly symmetric, consistent with a relaxed outer HI disk (\cite{mvg98}). Alternatively, while fly-by interactions can occur even in low density environments (e.g., \cite{kre11}), the known companions of NGC 5523 have projected separations of $>$1 Mpc and are unlikely to have made fly-bys past NGC 5523 given their relative velocities (Table 6). We therefore favor a past merger as the perturbation that produced a long-lived asymmetry in NGC 5523 (\cite{scj07}), an event that could also be associated with the production of its prominent bulge (\cite{bjc05}).

We infer that the nucleated bulge, the probable current dynamical center of the system, has always been the dynamical center of the primary host galaxy. There was likely a past perturbation that caused the entire inner disk to shift in comparison to this dynamical center, thus creating an offset (e.g., \cite{mng02}; \cite{bjc05}). This perturbation could have been caused by the object that is now the tidal stream if it were sufficiently massive, or more likely by an object that has been assimilated into the galaxy. Following this prediction, the brightness minimum is a direct result of the inner disk shifting and drawing away from the nucleated bulge. It does not represent the true center of the primary host galaxy, even though it is currently the isophotal center of the inner disk.

Such a significant perturbation throughout the inner disk of NGC 5523 could be explained by one of two scenarios. Both cases are interpreted in the context of the \cite{par16} simulations of galaxy interactions which are consistent with other models in showing that mergers and interactions can lead to non-axisymmetric galaxy structures (e.g., \cite{bjc05}; \cite{jnm06}). In the first scenario, a companion body of roughly 10$\%$ the mass of the primary galaxy merged with and has already been assimilated into the host galaxy. In this case, there is no direct evidence of the companion body, and the tidal stream is a remnant from a second non-disruptive satellite merger that occurred in the more-recent past.

Alternatively, the tidal stream is the remnant of the initial significant merger that caused the inner disk to shift. The tidal stream, as currently observed, is far too small to have created such an impact on the host galaxy, having a mass of only $\sim$0.1$\%$ of its host. Thus, in order for the object that is now the tidal stream to cause such a disturbance, it would have to have infiltrated the host galaxy along the direction of rotation or pass completely through the plane of the galaxy. We support the former scenario, that the object was assimilated along the direction of rotation, because the tidal stream is currently crossing the plane of the inner disk. In general, the presence of the tidal stream as a merging body within NGC 5523 suggests the presence of multiple dwarf satellites surrounding NGC 5523.

The highly regular outer disk of NGC 5523 offers insight into the timescale of this soft interaction. Given that a galaxy's structure will return to homogeny over time (\cite{znr97}; \cite{scj07}; \cite{lot15}), the extreme regularity and symmetry of the outer disk suggests that any mergers that NGC 5523 experienced occurred in the distant ($\geq$ few Gyr) past. Therefore, the disturbance must have been mild enough to maintain the integrity of the primary host galaxy (\cite{hop09}). We describe this type of interaction as a soft merger.

\section{Conclusion}

Our study of the evolution of the isolated galaxy NGC 5523 offers a counter example to simple assumptions about isolated galaxy evolution. The extreme isolation of NGC 5523 suggests that past mergers did not influence its evolution. However, its symmetric HI outer disk in combination with its asymmetric inner structure gives significant evidence for a past merger. Furthermore, the tidal stream, likely a remnant from an interaction with a bound dwarf companion, indicates a more-recent merger. We conclude that the asymmetric internal structure of NGC 5523 is the product of one or more past minor mergers with surrounding low-mass galaxies. 

We reach this conclusion based on the observed structural effects of interactions in NGC 5523 in combination with the $\geq$ Gyr timescales necessary for these effects to rehomogenize among the galaxy. Evidence for such interactions includes the offset between the inner/outer disks, the prominence of the tidal stream, and the properties of the nucleated bulge. NGC 5523's complex inner structure in combination with its global isolation and homogenous outer structure have most likely been produced by mergers with its nearest neighbors: a process of isolation by annexation.

\begin{acknowledgements}
We would like to thank the staff of the WIYN Observatory for their assistance in obtaining the images in this study. We thank Zishan Xia for early discussions and and Elena D'Onghia and Stephen Pardy for sharing their results on interacting galaxies prior to publication. Ralf Kotulla gratefully acknowledges financial support from the National Science Foundation under Grant No. AST-1412449. We would like to thank the entire ALFALFA survey team for their contributions towards observing, data processing and analysis. The ALFALFA survey team at Cornell has been supported by US NSF grants AST-0607007 and AST-1107390 to RG and MPH and by continuing support from the Brinson Foundation. This research has made use of the NASA/IPAC Extragalactic Database (NED) which is operated by the Jet Propulsion Laboratory, California Institute of Technology, under contract with the National Aeronautics and Space Administration. STSDAS is a product of the Space Telescope Science Institute, which is operated by AURA for NASA. Funding for the Sloan Digital Sky Survey IV was provided by
the Alfred P. Sloan Foundation, the U.S. Department of Energy Office of
Science, and the Participating Institutions. SDSS-IV acknowledges
support and resources from the Center for High-Performance Computing at
the University of Utah. The SDSS web site is www.sdss.org.

SDSS-IV is managed by the Astrophysical Research Consortium for the 
Participating Institutions of the SDSS Collaboration including the 
Brazilian Participation Group, the Carnegie Institution for Science, 
Carnegie Mellon University, the Chilean Participation Group, the French Participation Group, Harvard-Smithsonian Center for Astrophysics, 
Instituto de Astrof\'isica de Canarias, The Johns Hopkins University, 
Kavli Institute for the Physics and Mathematics of the Universe (IPMU) / 
University of Tokyo, Lawrence Berkeley National Laboratory, 
Leibniz Institut f\"ur Astrophysik Potsdam (AIP),  
Max-Planck-Institut f\"ur Astronomie (MPIA Heidelberg), 
Max-Planck-Institut f\"ur Astrophysik (MPA Garching), 
Max-Planck-Institut f\"ur Extraterrestrische Physik (MPE), 
National Astronomical Observatory of China, New Mexico State University, 
New York University, University of Notre Dame, 
Observat\'ario Nacional / MCTI, The Ohio State University, 
Pennsylvania State University, Shanghai Astronomical Observatory, 
United Kingdom Participation Group,
Universidad Nacional Aut\'onoma de M\'exico, University of Arizona, 
University of Colorado Boulder, University of Oxford, University of Portsmouth, 
University of Utah, University of Virginia, University of Washington, University of Wisconsin, 
Vanderbilt University, and Yale University.
\end{acknowledgements}

% WARNING
%-------------------------------------------------------------------
% Please note that we have included the references to the file aa.dem in
% order to compile it, but we ask you to:
%
% - use BibTeX with the regular commands:
%   \bibliographystyle{aa} % style aa.bst
%   \bibliography{Yourfile} % your references Yourfile.bib
%
% - join the .bib files when you upload your source files
%-------------------------------------------------------------------

\end{document}